# Improvement of Methods for Determination of Reological Parameters of Non-Newtonian Liquids.

N.G.Domostroeva[*] and N.N.Trunov[†]

*D.I.Mendeleyev Institute for Metrology*

*Russia, St.Peterburg. 190005 Moskovsky pr. 19*

(Dated: November 26, 2008)

**Abstract**: Methods for simple and effective determination of reological parameters of non-Newtonian liquids are proposed.

PACS numbers: 47.50.-d.47.15.Cb

Non-Newtonian liquids form a very wide, diverse and important set of media. Determining of the effective viscosity and other parameters of such liquids is connected with some inverse problem. Its practical solution may often require suitable methods and simple practical approximations.

First of all we consider a coaxial viscosimeter with radii $R_1$ and $R_2$, so that $a = (R_1/R_2)^2 < 1$. The external cylinder rotates with the angular velocity $\omega$, and the inner one is motionless. Then the gradient of the velocity $\gamma$ is connected with $\omega$ by means of the following equation

$$\omega(\tau) = \frac{1}{2}\int_{a\tau}^{\tau} \frac{\gamma(t)}{t}dt \qquad (1)$$

with $\tau$ being the tangential stress [1]. Introducing an auxiliary function

[*] Electronic address: N.G.Domostroeva@vniim.ru
[†] Electronic address: trunov@vniim.ru

$$\varphi(t) = \frac{\gamma(t)}{2t} \tag{2}$$

one can get from (1)

$$\varphi(t) = \frac{d}{dt}\sum_{k=0}^{\infty}\omega(ta^k) \tag{3}$$

This formula contains the differentiation and the infinite series. Besides an approximate calculation of the series (2) leads to the derivatives of higher order. Meanwhile practically $\omega(t)$ is only determined in few points. That's why difficulties may arise when using (3).

We propose to use for transformation (3) the Abel-Plana's summation formula whish has been successfully used for many problems [2]

$$\sum_{k=0}^{\infty} F(k) = \int_0^{\infty} F(x)dx + \frac{F(0)}{2} + \Delta[F] \tag{4}$$

$$\Delta[F] = i\int_0^{\infty}\frac{F(iy) - F(-iy)}{\exp(2\pi y) - 1}dy \tag{5}$$

Substituting in (4), (5) our

$$F(k) = \frac{d}{dt}\omega(te^{-kL}), \qquad L = |\ln a| \tag{6}$$

we obtain

$$\varphi(\tau) = \frac{\omega(\tau)}{\tau L} + \frac{\omega'(\tau)}{2} + \Delta \tag{7}$$

with $\Delta \square L$. Here small parameter is $L = |\ln a|$ instead of the usual $1-a$.

A frequent supposition is that $\omega(\tau) = A\tau^{\beta}$. Then from (1), (2) we get immediately

$$\varphi(\tau) = \frac{\beta A}{1 - a^{\beta}}\tau^{\beta - 1} \tag{8}$$

The same result follows certainly from (4)-(7) but it requires the explicit calculation of $\Delta$, since we use here the parameter L and no $a$ immediately. There are several ways to get suitable simple and practically exact enough formulas for the functions $\omega(\tau)$ close to the power-law ones. First of all, we can use in (6) an explicit form of $\omega$ and $\omega'$ but calculate $\Delta$ for some average suitable value of $\beta$:

$$\Delta = At^{\beta-1}\left[\frac{\beta}{2th(\beta/2)} - \frac{1}{L}\right] \qquad (9)$$

Since (7) is really a power series expansion in the usually small parameter $L$ and $\Delta \square L^2$, such replacement of the exact $\Delta$ causes only small deviation of $\varphi(\tau)$.

Moreover the Abel-Plana's formula may be used as a base for various approximations; in particular for a very effective two-point Pade approximation

$$\omega(t) = At^{\beta}\frac{1+Bt}{1+Ct} \qquad (10)$$

One can offer several simple calculations for approximately power-law function $\omega(\tau)$ as

$$\omega(\tau) = A\tau^{\beta}\left(1 + B\tau + C\tau^2 + ...\right) \qquad (11)$$

with small coefficients B, C. From (1), (2) we get

$$\varphi(\tau) = \frac{\beta A\tau^{\beta-1}}{1-a^{\beta}}\left[1 + D\tau + E\tau^2\right] \qquad (12)$$

with small

$$D = \frac{\beta+1}{\beta}B\frac{1-a^{\beta}}{1-a^{\beta+1}}$$
$$E = \frac{\beta+2}{\beta}C\frac{1-a^{\beta}}{1-a^{\beta+2}} \qquad (13)$$

Another way is to introduce an effective exponent

$$b(\tau) = \frac{\omega(\tau)}{\int_0^{\tau}\frac{\omega(\tau)dr}{\tau}} \qquad (14)$$

which coincides with $\beta$ in the case $\omega = A\tau^{\beta}$. Then an interpolation formula

$$\varphi(\tau) = \frac{\omega(\tau)}{\tau}\frac{b(\tau)}{1-e^{-bL}} \qquad (15)$$

is exact in two limited cases: 1) $L \to 0$, i.e. $a \to 1$ and an arbitrary function $\omega(\tau)$; 2) arbitrary value of $L$ and the power-law $\omega(\tau)$. It allows to suppose that (15) is exact enough in intermediate cases.

For viscosimeters consisting of two cones the corresponding equation is more complicated

$$\omega(\tau) = \int_{c\tau}^{\tau} \varphi(t) \sqrt{\frac{t}{t-bt}} dt \qquad (16)$$

$$0 < c < 1, \qquad b \le c$$

where we denote $b = \sin^2 \alpha_1$, $c = b/\sin^2 \alpha_2$ and $\alpha_1 < \alpha_2 \le \pi/2$ are two angular openings of the cones. In the limiting case $\alpha_2 = \pi/2$ we have a plane-cone viscosimeter with $c=b$. It is convenient to factorize the dependence of $\omega$ upon $\varphi$ and upon parameters $b, c$ of two cones. We introduce a new variable $z$ so that $t = z\tau$ and use it in the Mellin transforms

$$\Phi(s) = \int_0^\infty \varphi(t) t^{s-1} dt \qquad (17)$$

$$\Psi(s) = \int_0^\infty \frac{\omega(t)}{t} t^{s-1} dt \qquad (18)$$

for $\varphi$ and $\omega/\tau$ respectively. Thus we get

$$\Psi(s) = A(1-s,c,b)\Phi(s) \qquad (19)$$

$$A(x,c,b) = \int_c^1 \sqrt{\frac{z}{z-b}} z^{x-1} dz \qquad (20)$$

so that all parameters of our viscosimeter are concentrated in the factor $A(s,c,b)$. For a given $\omega$ we get $\Psi$ and then $\Phi$ from (19). The reverse Mellin transformation accordingly to the known formulas or tables leads to $\varphi(\tau)$ sought. For complicated forms of $\omega$ some approximations may be useful or unavoidable.

A lot of calculations are made for supposed function $\omega = At^\beta$. So it is reasonable to study a close form

$$\omega(\tau) = B\tau^\beta \sum_{k=0}^\infty p_k t^k \qquad (21)$$

with $p_o = 1$ and relatively small coefficients $p_k, k \ge 1$. Such form must be suitable for a wide class of fluids. Using the above method we get

$$\varphi(\tau) = \frac{B\tau^{\beta-1}}{A(\beta,c,b)} \sum_{k=0}^\infty p_k s_k(\beta,c,b) \tau^k \quad, \qquad (22)$$

$$s_k = A(\beta,c,b)/A(\beta+k,c,b)$$

so that $p_0 s_o = 1$. All the coefficients $s_k$ are of the order unity, at last for moderate values of $k$. Practically we can put $p_k \neq 0$ only for small values of $k$.

All the above methods can also be adapted and used for another types of non-Newtonian liquids and for studying the Weissenberg effect.